# On the importance of Internet eXchange Points for today's Internet ecosystem

Nikolaos Chatzis
TU Berlin
nikolaos@net.t-labs.tu-berlin.de

Georgios Smaragdakis
T-Labs/TU Berlin
georgios@net.t-labs.tu-berlin.de

Anja Feldmann
TU Berlin
anja@net.t-labs.tu-berlin.de

## Abstract

Internet eXchange Points (IXPs) are generally considered to be the successors of the four Network Access Points that were mandated as part of the decommissioning of the NSFNET in 1994/95 to facilitate the transition from the NSFNET to the "public Internet" as we know it today. While this popular view does not tell the whole story behind the early beginnings of IXPs, what is true is that since around 1994, the number of operational IXPs worldwide has grown to more than 300 (as of May 2013[1]), with the largest IXPs handling daily traffic volumes comparable to those carried by the largest Tier-1 ISPs, but IXPs have never really attracted any attention from the networking research community. At first glance, this lack of interest seems understandable as IXPs have apparently little to do with current "hot" topic areas such as data centers and cloud services or software defined networking (SDN) and mobile communication. However, we argue in this article that, in fact, IXPs are all about data centers and cloud services and even SDN and mobile communication and should be of great interest to networking researchers interested in understanding the current and future Internet ecosystem. To this end, we survey the existing but largely unknown sources of publicly available information about IXPs to describe their basic technical and operational aspects and highlight the critical differences among the various IXPs in the different regions of the world, especially in Europe and North America. More importantly, we illustrate the important role that IXPs play in today's Internet ecosystem and discuss how IXP-driven innovation in Europe is shaping and redefining the Internet marketplace, not only in Europe but increasingly so around the world.

## Categories and Subject Descriptors

A.1 [**Introductory and Survey**]; C.2.1 [**Network Architecture and Design**]; C.2.3 [**Network Operations**]: Network Management; C.2.6 [**Internetworking**]: General

---

[1] [25] lists 391 as of May 2013, but includes also some IXPs that were not up and running (i.e., fully operational) as of May 2013.

## 1. INTRODUCTION

Recent findings and events concerning different aspects of the Internet ecosystem have been instrumental in bringing Internet eXchange Points (IXPs) to the attention of network researchers, have heightened the concern among network operators with respect to the IXPs' proper operations, and have raised awareness about IXPs among the public in general. First, for networking researchers, the 2012 study by Ager et al. [43] that analyzed sFlow measurements from a large European IXP and revealed the presence of some 50K+ actively used peering links in this single IXP was an eye-opener. To put the discovery of such a rich peering fabric supported at this single IXP into proper perspective, recall that for many years, the networking research community has been studying AS-level maps of the global Internet that had far fewer peering links *in total* than was observed in just this single location. In the process, the authors of [43] also highlighted the increasingly important role that such large IXPs play especially in Europe, where their daily traffic volumes are comparable to those seen by the largest global Tier-1 ISPs, where their constantly growing list of member ASes reads like the "who is who in Internet commerce" and where their service offerings rival and compete head-on with those traditionally provided by the incumbent ISPs.

Second, the critical role of network operators to ensure "best-in-class" operations of IXPs and a clear need for networking researchers to know more about these IXPs' basic mode of operations and underlying business models have come to the forefront in the recent discussions in the trade press [18, 5, 13] and technology blogs [35, 33, 36, 6] concerning the highly publicized DDoS attacks in late March 2013 featuring players like Spamhaus, Stophaus, CloudFlare, and IXPs [28]. In short, in an attempt to dilute the massive DDoS attacks launched by Stophaus against Spamhaus, Spamhaus apparently started to rely on CloudFlare's services and remained accessible. To counter Spamhaus' tactic, Stophaus allegedly redirected its effort and started to DDoS the network components that are at the core of CloudFlare's technology: the IXPs where CloudFlare connects to its providers for the purpose of exchanging traffic. Irrespective of the motives of the involved parties or the accuracy of the reported actions, these and similar incidents demonstrate that understanding the end-to-end flow of traffic in today's Internet has to include as key component the IXPs and the networks that peer at those IXPs.

Lastly, given the widely acknowledged differences between the European IXP scene and the IXP marketplace in North America, very recently announced intentions [67, 52, 40] to revive the relatively stagnant North-American IXP marketplace by trying to gain a foothold for the European exchange model in the US and Canada could have far-reaching implications for a wide range of Internet stakeholders. Most importantly, if successful, such an attempt would address the scarcity of public peering opportunities in North America in comparison to Europe and would clearly challenge the dominant commercial (i.e., for-profit) IXP business model in the US. In view of such a potential business model "import" from Europe to North America, it will be important to understand the key reasons for the enormous attraction that these large IXPs have in Europe and examine the differences in economic conditions that may or may not favor a status quo over a complete overhaul for the North American IXP scene.

Motivated by these and other examples, this article serves on the one hand as a short tutorial where we describe the basic operations of a generic IXP and the typical set of service offerings that networks can use once they become a participant of such an IXP. Given the significant differences in the various IXP marketplaces, we also explain the key aspects that distinguish the European IXP scene from the North American one or from IXPs in the Asian-Pacific, African, or Central/South-American regions. All the information provided in this article is publicly available and may be well-known within the NANOG/network operator community. However, the critical role that IXPs are playing in the overall Internet ecosystem has until recently gone unnoticed by large parts of the networking research community, and much of the provided background material may be essentially unknown to many networking researchers whose focus has traditionally been US-centric, meaning that their emphasis to date has been typically on large ISPs, content delivery networks (CDNs), and large content providers and not on what appears at first glance relatively unimportant components of the Internet infrastructure in the form of US-based IXPs.[2] In this sense, this tutorial is also intended to be a survey of the existing literature on IXP-related and relevant works that have not necessarily appeared in mainstream networking research venues.

At the same time, we also view this article as an opportunity to showcase IXPs as a rich source for interesting networking research problems that have largely gone unnoticed in the past. For one, IXPs are essential for studying the Internet ecosystem, including the increasingly sophisticated peering strategies deployed by many of today's Internet players and the increasingly complex nature of the traffic flow across the network that results from these peering arrangements. In particular, having a solid grasp of the role that IXPs play in the existing Internet ecosystem is critical for understanding how content is distributed in today's Internet and how the different parties (e.g., content providers, CDNs, ISPs) are adapting to the constantly changing nature of content distribution. Moreover, IXPs are bound to play a similarly important if not more critical role for the emerging cast of cloud providers and their attempts to bring their services as close to the end users as possible. Finally, from a more IXP-centric perspective, the sustained rapid growth in IXP members as well as IXP traffic volumes in some of the most successful IXPs combined with the demand for and need to properly manage increasingly complex IXP-specific traffic engineering solutions raises serious scalability issues with respect to IXPs' switching architectures and network operations and may provide concrete application domains for Software Defined Networking (SDN) and also inspire new research problems in that space.

Finally, while this article is intended for the general networking research community, it is not meant to and cannot subsume the excellent expositions on a wide range of IXP-related issues by W. Norton (e.g., see [58]). On the contrary, we strongly encourage the readers of this article to use it as a stepping stone towards becoming more familiar with the many related topics discussed by Norton and incorporating many of his more economics-focused arguments into future studies of the Internet ecosystem. While IXPs have occasionally featured in networking research papers (see, for example, [48, 69, 53, 44, 63, 47, 65, 51], to mention but a few), they are typically the focal point of Norton's white papers on all aspects related to peering (e.g., [57, 60, 59]) and of reports by organizations such as Euro-IX [64, 16], Internet Society [54, 55], or governing or regulatory bodies such as OECD [21, 68], EU [29], FCC [42, 38], and ITU [41, 39]. Our hope is that this article will raise awareness among networking researchers about the importance and relevance of IXPs for today's Internet by connecting ongoing or planned research efforts in this area more tightly with the expertise that exists within the network operator/provider communities on IXPs and IXP-related issues but is often ignored or not understood by the more research-oriented networking community.

## 2. IXP 101

Although IXPs come in many different shapes, shades and sizes, some aspects concerning their architecture, operations, and service offerings are somewhat generic and are applicable to most of the existing operational IXPs across the globe with some legitimate business focus.

### 2.1 Historical perspective and notes

The decommissioning of the National Science Foundation Network (NSFNET) around 1994/95 refers to a carefully orchestrated plan for transitioning the NSFNET backbone service to private industry. That plan involved the establishment and operation of a set of four Network Access Points (NAPs) located in the US (i.e., Washington, D.C., New York/New Jersey, Chicago, and California); that is, public network exchange facilities where commercial Internet Service Providers (ISPs) connected with one another to exchange traffic.[3] As a transitional strategy, these facilities were effective and partly responsible for a smooth transition from a largely monolithic network that started as a government-funded academic experiment to what marked the beginnings of the modern Internet – a network of networks comprised of an increasingly diverse set of players that interconnect with one another to exchange reachability information for the ultimate purpose of doing business or selling their services to end users or other customers. Since their creation around 1995, these four government-mandated NAPs in North America have been replaced by some 80 modern IXPs (86 as of early 2013) and can be found in all major cities or metropolitan areas. They are typically operated as for-profit commercial entities[4] and continue to fulfill by and large the NAPs' original role of providing a physical infrastructure and carrier-neutral meeting place where ISPs or other networks can communicate with each other (i.e., exchange their traffic locally), independent of third parties.[5] Interestingly, some of the largest players in the North American IXP marketplace don't market themselves primarily as IXPs but as pioneers in data center and interconnection services (e.g., Equinix,

---

[2]For notable exceptions, see [56] who tracked the Internet's early infrastructure from 1997-2000 as well as recent studies for policymakers [62, 68].

[3]NAP-like facilities existed before the decommissioning of the NSFNET in the form of the Commercial Internet eXchange (CIX) on the West coast of the U.S., the Metropolitan Area Ethernet (MAE-East) on the East coast, and the two Federal Internet eXchanges FIX-E and FIX-W [17].

[4]However, there are exceptions such as the non-profit Seattle Internet Exchange SIX that operates according to a rather unique donor model (i.e., there is low/no cost for participant networks and no fees) [32].

[5]Although the original NAPs have been replaced by modern IXPs, in Spanish-speaking Latin America, the term NAP lives on in IXP names like "NAP do Brazil".

Telehouse America, telx) and are less than forthcoming about their network infrastructures, operations, and business strategies.

In the early 1990s[6], NAP-like facilities also sprang up in Europe (e.g., London, Frankfurt, Amsterdam), but for very different reasons than in the US. As commercial Internet traffic increased rapidly in Europe due to faster Internet access, many of the incumbent Telcos turned ISPs and many of their new competitors that appeared in the form of regional ISPs recognized early on that linking their networks for the purpose of exchanging their local traffic locally would be beneficial for all involved parties because it would avoid paying the astronomical transatlantic bandwidth costs. Driven in large parts by such purely economic considerations, the European IXP marketplace has strived from the very beginning but evolved very differently from its North American counterpart. In particular, the not-for profit IXP business model that was adopted in much of Europe from the get-go has flourished as the Internet has grown by leaps and bounds by any imaginable metric during the past 15-20 years. This business model has contributed to an extremely vibrant and innovative European IXP scene that consists today of some 150 IXPs and represents an impressive spectrum of players, ranging from the largest IXPs worldwide (i.e., AMS-IX, DE-CIX, LINX) to up-and coming IXPs (e.g., MSK-IX) and critical regional players (e.g., Netnod, UA-IX, ESPANIX, PLIX, France-IX, ECIX, VIX) all the way to small local IXPs that can be found all across Europe. In fact, some of the largest European IXPs are comparable to the largest global Tier-1 ISPs in terms of the amount of traffic they handle on a daily basis, the number of networks that are responsible for that traffic, or the number of countries from where that traffic originates from or is destined to. Most of these IXPs are very open about most aspects of their business and typically maintain websites with detailed technical specification of their infrastructure, up-to-date membership lists, aggregate traffic statistics, and specifics of their service offerings.

## 2.2 Basic operations and services

To start, according to a commonly-used definition, an IXP is *a network infrastructure with the purpose to facilitate the exchange of Internet traffic between Autonomous Systems (ASes) and operating below layer 3. The number of ASes connected should at least be three and there must be a clear and open policy for others to join* [16].

This or similar definitions allow for a flexible interpretation of the types of network infrastructures that can be considered as IXPs. For example, the infrastructure can be something as simple as a single switch in a basement, or as complex as the planned DE-CIX Apollon platform whose core will be distributed across four supernodes located at highly secure and resilient data centers in Frankfurt, will utilize ADVA's flagship FSP 3000 scalable optical transport solution, and will be built on a switching layer supported by Alcatel-Lucent's 7950 Extensible Routing System. Figure 1 depicts the current (from 2012) topology of DEC-IX and clearly shows the built-in redundancy at the different aggregation levels (e.g., DE-CIX3 and DE-CIX6 represent the redundant core of the switch fabric) and also the the distributed nature of the switch (i.e., DE-CIX1-4 and DE-CIX7 are the five collocation sites of DE-CIX within the city of Frankfurt).

An IXP's entire infrastructure can be located in a single physical facility (e.g., the Cairo Internet exchange CAIX at 26 Ramses Street, in Cairo, Egypt), in multiple locations within the same city (e.g., DE-CIX) or region (e.g., ECIX), or it can be distributed at global scale (e.g., the Equinix Internet Exchange is distributed across 19 locations in 17 global metro areas). Yet other IXPs consist of multiple geographically dispersed not-interconnected network infrastructures (i.e., an "IXP of IXPs"; e.g., Netnod in Scandinavia, or AMS-IX in Amsterdam AMS-IX and its "branch" AMS-IX HK in Hong Kong) and use different business solutions for interconnecting the different pieces. For example, while in the case of Netnod, the inter-IXP connectivity is provided by the larger Swedish ISPs that are participants in all of Netnod's IXPs, AMS-IX negotiated an exclusive arrangement with Hutchison Global Communication (HGC) to provide connectivity between Amsterdam and Hong Kong.[7] However, irrespective of the their size and architecture, IXPs with a commercial bent typically deploy a fully redundant switching fabric to provide an extra level of fault-tolerance and house their equipment in facilities such as data centers that are known for high levels of reliability (e.g., full UPS power backup, 99.999999% uptime), power density (e.g., heating, ventilation, AC), and security (e.g., facility access control and monitoring).[8]

---

[6]The CERN Internet eXchange Point CIXP in Geneva was, in fact, established in the late 1980s.

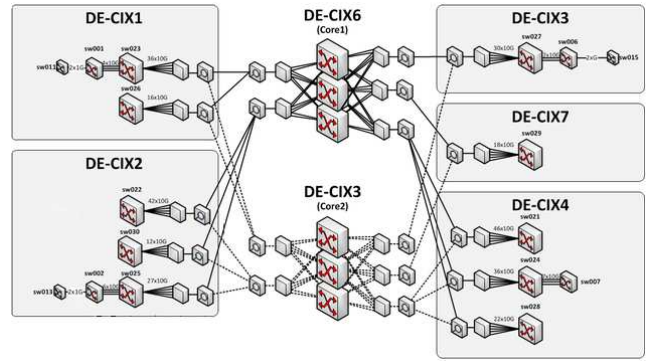

**Figure 1: Example of an IXP topology: DE-CIX in Frankfurt (circa 2012).**

With the vast majority of today's operational IXPs relying on an Ethernet-based switching fabric, networks (also called participants) that want to exchange traffic at a given IXP are generally expected to comply with the following basic requirements:

1. Each participating network must have a public Autonomous System Number (ASN).

2. Each participant brings a router to the IXP facility (or one of its locations in case the IXP has an infrastructure distributed across multiple data centers) and connects one of its Ethernet ports to the IXP switch and one of its WAN ports to the WAN media leading back to the participant's network.

3. The router of each participant must be able to run BGP since the exchange of routes across the IXP is via BGP only.

4. Each participant has to agree to the IXP's General Terms and Conditions (GTC).

Thus, for two networks to *publicly peer* at an IXP (i.e., use the IXP's network infrastructure to establish a connection for exchang-ing traffic according to their own requirements and business relationships), they each incur a one-time cost for establishing a circuit from their premises to the IXP, a monthly charge for using a chosen IXP port (higher port speeds are more expensive), and possibly an annual fee for membership in the entity that owns and operates the IXP. In particular, exchanging traffic over an established public peering link at an IXP is in principle "settlement-free" (i.e., involves no from of payment between the two parties) as IXPs typically do not charge for exchanged traffic volume. Moreover, IXPs typically do not interfere with the bilateral relationships that exists between the IXP's participants, unless they are in violation of the GTC. For example, the two parties of an existing IXP peering link are free to use that link in ways that involve the transfer of money from one party to the other (i.e., "paid" peering), and some networks may even offer transit across an IXP's switching fabric. Depending on the IXP, the time it takes to establish a public peering link can range from a few days to a couple of weeks.

---

[7]Yet other examples of an operational "IXP of IXPs" are Lyonix/Topix interconnecting the Lyon (France) and Turin (Italy) regions, BalkanIX (Sofia, Bulgaria) and InterLAN (Bucarest, Romania), or France-IX and SFINX (both in Paris, France) [10].

[8]As an illustration of the resilience of modern IXPs, LINX reported its first significant outage in 15 months on 5/31/2012 [28] and a recent report [66] highlights their importance to national cyber-defense.

By providing a well-defined set of steps and requirements for establishing public peering links between participating networks, IXPs clearly satisfy the main reason for why they exist in the first place – keeping local traffic local. However, there are other compelling reasons for why networks may want to connect to an IXP. For one, an IXP's public peering service is typically offered at a cost that is below the cost incurred by exchanging that same traffic using more conventional means; that is, relying on third-parties (i.e., the participants' upstream providers) to handle the traffic, typically for a traffic volume-related price. In addition, IXP participants often also experience improved network performance and QoS due to reduced delay (e.g., decreased round-trip times) and routing efficiencies (e.g., reduced number of AS hops for typical end-to-end paths). Last but not least, companies that are critical players in today's Internet ecosystem (e.g., Google) often "incentivize" other networks to connect at IXPs by making a network's presence at a certain IXP or certain number of IXPs an explicitly stated requirement for peering with them [20, 23].

In theory, with such a compelling list of reasons for networks to become participants at IXPs and peer there with selective partner networks, IXPs are naturally expected to further increase their attractiveness to all types of networks by offering additional services, either for free or for a price. For instance, given an already existing set of participating networks, it is not surprising to see that most operational IXPs also have offerings for "private" peering; that is, Private Interconnects (PI) that do not use an IXP's public peering infrastructure and enable direct traffic exchange between the two parties of a PI. In contrast to public peering, private peering is typically used between two IXP participants that are interested in having a well-provisioned dedicated link that can handle relatively stable but generally high-volume bi-directional traffic.

Other service offerings that are provided by an increasing number of today's IXPs include service level agreements (SLA)s for participants and the free use of the IXP's route-server(s). By offering the use of its route server(s), an IXP enables its participants to arrange instant peering with a large number of co-located participant networks using essentially a single agreement/BGP session (we return to this innocuous-looking service and discuss its importance in Section 3 below). Yet other service offerings are designed to attract new networks by making access to the IXP easier. A particularly popular such offering involves IXP reseller or partner programs that allow third parties (i.e., resellers such as IX Reach or Atrato) to resell IXP ports anywhere where they have infrastructure connected to the IXP. Under such programs, these third-parties and their services are certified to offer the IXP's service remotely which, in turn, enables networks that generate/attract little traffic to nevertheless use the IXP. Similarly, it also allows networks that serve regions in distant geographic areas from the IXP to connect to the IXP (also known as "remote peering"). Newer services that are provided by selective IXPs are aimed at particular providers and include, for example, support for mobile peering (i.e., a full-service, scalable solution for the interconnection of mobile GPRS/3G networks) and customer-triggered blackholing to help participants to mitigate the effects of Distributed Denial of Service (DDoS) attacks against their networks. Lastly, as part of their mission to work for the "good of the Internet", some IXPs (e.g., the Scandinavian IXP Netnod) offer additional free value-added services such as Internet Routing Registry (IRR), consumer broadband speed tests[9], DNS root nameservers, country-code top-level domain (ccTLD) nameservers, as well as distribution of the official local time through NTP.

## 2.3 Business and operational models

In addition to local economic conditions and overall market conditions, the size or geographic reach of an IXP as well as its range of service offerings are ultimately a function of the IXP's business and operational model – who runs and operates the IXP, for what reasons/purpose, and with what vision or goals in mind. To this end, we distinguish to a first approximation between "for-profit" and "non-profit" IXPs, and subdivide the latter further into "cooperative" and "managed" non-profit IXPs. One of the most striking observation when looking at the worldwide IXP marketplace is that outwardly, the largest and most vibrant and innovative IXPs reside in Europe and are managed non-profit IXPs (e.g., DE-CIX, AMS-IX, LINX). For example, DE-CIX is wholly owned by the eco association, the world's largest non-profit association for the Internet industry. As such, DE-CIX views its participant networks as customers (or participants), not as members or shareholders. The IXP's management team reports to the eco association and not to its own customers, though they can provide input into the operation and management of the IXP through an advisory board. Similar governance structures apply to AMS-IX and LINX and, like in the case of DE-CIX, have undoubtedly contributed to the enormous success of these IXPs have as worldwide leaders in the global IXP marketplace. Due to their openness and explicitly stated mission to work for "the good of the Internet", much is known and publicly available about these managed non-profit IXPs, including up-to-date topologies with detailed technical specifications and descriptions, expansion plans, lists of connected networks (customers), traffic statistics, service offerings and pricing lists, and outreach activities. Collectively, we refer to these managed non-profit IXPs as representing the "European IXP model".

In stark contrast to the European IXP scene, the North American IXP marketplace is dominated by for-profit IXPs whose first and foremost interest (indeed, mandate) is to generate profits for their shareholders. As such, these IXPs are much less forthcoming with providing a similar level of detailed information about their architectures, operations, and service offerings as their European counterparts, and as a result, our understanding of the North American IXP marketplace is more murky. What seems clear is that the largest IXPs in North America that publish their member lists and report traffic statistics are significantly smaller than their European non-profit counterparts with respect to pretty much any applicable metric. However, this picture does not account for companies such as Equinix or Telehouse America that market themselves first and foremost as pioneers in the deployment and management of data centers or collocation facilities on a global scale. While pro-viding IXP-like services is not necessarily their primary objective, advertised service offerings such as Equinix Internet Exchange or Global Interlink are nothing but globally distributed network infrastructures facilitating the interconnection of the companies' customers. For example, based on marketing brochures published by Equinix [27], the Equinix Internet Exchange aggregates thousands of peering sessions onto a shared fabric, connecting peers at 19 IXP locations in 17 global metro areas. With over 750 network, content, and cloud providers peering on the massively distributed global Equinix Internet Exchange platforms, Equinix calls itself "the Home of the Internet". However, when viewed as an IXP, its public peering fabric is generally believed to be no match for those supported by the largest European IXPs (the opposite may be true with respect to private peerings, though). We refer to Equinix and the other for-profit North American IXPs as following the "US IXP model".

---

[9]FCC also considers IXPs as candidate locations to assess QoS [42].

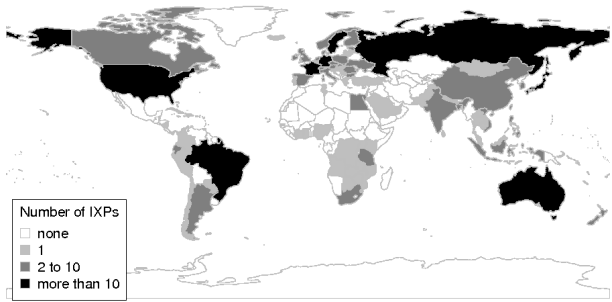

Figure 2: Number of IXPs per country (data from PCH).

As for the other IXP marketplaces around the world, the Asia/Pacific region consists of both for-profit and non-profit IXPs, where the former are more likely to be found in the more developed countries and the latter are typically encountered in the less developed countries [62]. Latin American IXPs are almost exclusively non-profit, and so are all currently operating IXPs in Africa. In general, different attempts over time by local governments or regulators in various places around the world to mandate how IXPs should operate (e.g., require "forced" peering; that is, each participant has to peer with each other participant at the IXP) have done little to encourage a more active IXP marketplace. At the same time, there have been instances where the involvement of local governments was critical to fight the legal maneuvers of strong–willed incumbent Telcos (e.g., Kenya) and ensure the continued operation of IXPs.

## 3. THERE IS MORE TO IXPS THAN MEETS THE EYE

By delving deeper into what kind of information about IXPs is generally available and how IXPs operate as part of the Internet's ecosystem, we next highlight some aspects of IXPs that explain to some extent their success and popularity in some parts of the world.

### 3.1 Know your IXP data sources

When it comes to studying the worldwide IXP scene more quantitatively, a number of well-known publicly available resources exist, including the widely-used database PeeringDB [26]; a often-cited IXP directory with associated meta-data maintained by Packet Clearing House (PCH) [25]; detailed lists, repositories, and reports provided as a public service by non-profit organizations such as the European Internet Exchange Association (Euro-IX) [15]; an IXP database from Data Center Map [7], a free web service acting as a link between providers and clients in the data center industry, etc. While these and similar resources are invaluable and provide often detailed information about many IXP-related aspects, their use for quantitative scientific studies requires great care because the voluntary nature of most of these data gathering efforts implies that the accuracy and freshness of the collected data is largely unknown. Even for PeeringDB which is often referred to as the industry database for peering information for network operators [20], this problem becomes obvious in cases where a comparison of the available information against ground truth (i.e., IXP-proprietary measurements that reflect actual traffic, peerings, policies, etc.) is possible [43]. What complicates any quantitative studies even further is the fact that many of the lesser-known IXP databases rely on entries from "more primary" sources (e.g., PeeringDB) to partly populate their repositories. For another example, it is the stated policy of PCH to never drop an IXP from its list; instead, only when sufficient evidence exists (e.g., through direct contacts), IXPs are marked as "defunct" or "down". As a direct result of this policy, the often-quoted PCH number of about 391 IXPs worldwide (as of May 2013) does not refer to the number of IXPs that were up and running (i.e., fully operational) as of May 2013. The number of IXPs per country as reported by PCH [25] is shown in Figure 2.

As a general rule, the managed non-profit IXPs provide on their websites the most accurate, reliable, and up-to-date information. Since the number of connected networks (participants) and traffic volumes handled by their infrastructures are often used as metrics for their success, it has become standard for those IXPs to announce each newly added participant and provide real-time aggregate traffic statistics. However, expect for some rare cases, even these managed non-profit IXPs are generally very protective of their customers' inherently more sensitive data related to their operational model and prevailing business relationships and publish neither the IXP peering matrix (i.e., which participant is publicly peering which other participants) nor its corresponding traffic matrix (i.e., how much hourly/daily traffic is exchanged on each of the established public peering link at the IXP). As a result, much of the information that networking researchers would be cherishing the most when studying the Internet ecosystem at the level of interconnected ASes (e.g., AS-level connectivity, AS-level traffic flows) has remained largely unknown and essentially impossible to be inferred from the widely-used public collections of BGP measurements (e.g., RouteViews [31], RIPE RIS [30]) or readily available datasets that have resulted from large-scale active measurement campaigns based on traceroute. In fact, due to their longstanding reputation of being extremely difficult to detect, AS-links of the peer-peer type representing public peerings at IXPs have been termed "invisible" links [61], and while their presence in the Internet has been well-known and generally acknowledged, what came as big surprise to most networking researchers was the discovery in [43] that a single IXP in Europe had more peering links than were assumed to exit Internet-wide.

### 3.2 Surprised by the obvious

The observation in [43] of an IXP with some 400 connected networks and a peering matrix with a fill rate of some 60-70% (i.e., some 50K+ active peerings out of some 80K possible peerings) was stunning as it turned the existing mental model of the AS-level Internet upside-down – not only did the number of peering links seen at this single IXP exceed the estimated number of all peering links in the entire Internet, but even by very conservative estimates, there are easily more than 200K public peering links in today's Internet, meaning that the ratio of AS links of customer-provider type to peer-peer type is not, as has been commonly assumed, about 3:1 but more like 1:2 or 1:3. The economic ramifications of this revised mental model should be obvious.

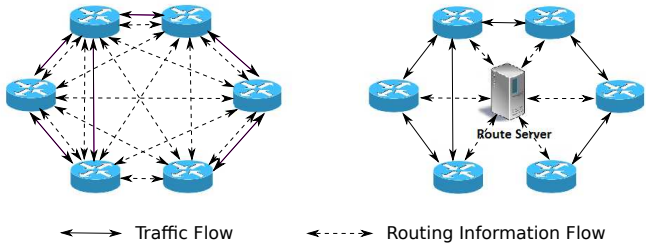

**Figure 3: Multi-lateral peering in an IXP without route server (left) vs. corresponding traffic and routing information flow in an IXP with a route server (right).**

At the same time, observing such a rich peering fabric supported by an actual IXP raises the question why this discovery took many networking researchers by surprise or, for that case, how such rich peering can occur in the first place. The answer to both questions is a perfect example of how little our community knew (or knows) about IXPs and is best phrased as another question: "Ever heard about route servers?" As already alluded to in Section 2.2, more and more IXPs, primarily in Europe, offer the free use of their route server as value-added service to their customers. By offering this free service, these IXPs greatly facilitate their customers' task of multi-lateral peering by drastically reducing the overhead and management complexity that would otherwise be required from each customer to establish and maintain BGP sessions between its router and each of its peers' routers. In short, a route server at an IXP is a process that (i) collects routing information from the (border) routers of the members that connect to it, (ii) typically does some processing on that information (e.g., route filtering), and (iii) distributes the (processed) information to the border routers of the members that use the route server. As a result, the routes that each of the IXP's members' border routers learns from the route server are the same as the routes each member would have learned from its direct peer. Thus, by using the IXP's route server, a member can replace some selectively-chosen or all of its BGP sessions with a single BGP session to the route server.[10] Figure 3 illustrates multi-lateral peering at an IXP without a route server (left) and with a route server (right) by depicting the corresponding information flows in the control and data planes.

The concept of operating a route server at an IXP and offering its free use to the IXP's members originated within the European IXP model and exemplifies its operational model. Clearly, as the membership of an IXP grows, the task of each individual member to establish bi-lateral peering agreements with all other interested parties at the IXP becomes time-consuming and cumbersome. Managed non-profit IXPs are a perfect vehicle for listening to their members' input and requests and coming up with a service that makes life for its members easier (e.g., operating a route server to facilitate multi-lateral peering for their members). Indeed, by providing the use of their router server as free value-added service, IXPs have made is extremely easy for new members to join – they can start exchanging routes almost immediately (i.e., once a BGP session with the route server is established), and traffic can start flowing from day one. In effect, the time-consuming negotiations and resulting contractual agreements of traditional peering agreements have been replaced by largely instantaneous informal handshake agreements between the two parties of a peering [21]. Similarly, the free route server offering has also been very beneficial for IXP members with an open peering policy (e.g., many CDNs), because for them, one BGP session suffices to peer with all other IXP members that have also have an open peering policy and use the IXP's route server. However, using the IXP route server has also benefits for members that do not have a open peering policy. For example, it allows them to establish direct BGP sessions with their most important peers only and rely on peering via the route server with all the other members, and it also provides a backup in case their direct BGP sessions become inactive. In terms of numbers, currently about 60-70% of all members of each of the largest European IXPs have an open peering policy and connect to the IXP's route server. The remaining 30-40% of the members connect directly with their most critical peers at the IXP. Thus, for an IXP with some 400 members, these numbers translate directly into approximately 50K+ peering links at such an IXP alone which fully explains the surprising result reported in [43].

### 3.3 Where non-profit meets for-profit

When examining in more detail the enormous growth that some of the large European IXPs have seen over the years in terms of number of connected networks and amount of traffic handled by their infrastructures (e.g., daily volumes, peak rates), some interesting patterns start to appear. For one, for IXPs such as DE-CIX or AMS-IX that are pushing towards the 500 and 600 number in terms of connected networks and multiple Tbps in terms of peak traffic rates, the uptake of some 10-20% new connected networks year after year [11], the demand for increasingly higher port speeds by these new or the existing customers (e.g., 100GE ports are the latest offerings [8, 14, 3]), and resulting sustained annual traffic growth rates of 50-100% entail an ever-expanding distributed network infrastructure. To illustrate, Table 1 shows the annual traffic and member statistics for AMS-IX for the last 10 years [2, 50]; the impressive traffic growth that AMS-IX has experienced in that same period is shown in Figure 4 where we plot the daily traffic volumes in petabytes (PB). from 2003 til early 2013. Some of the largest IXPs, such as AMS-IX in Amsterdam or DE-CIX in Frankfurt, reportedly carry on a daily basis similar amounts of traffic as some large ISPs (e.g., AT&T, Deutsche Telekom[11]).

Second, to accommodate such expansions, these IXPs that pride themselves to be strictly neutral (i.e., open to any network and independent of third-party companies, see also [37]) place equipment into new strategically located data centers within their geographic reach to offer new or improved access to their switching fabric. However, to maintain their neutrality, these non-profits strive for diversity without sacrificing quality and typically deploy in a number of different carrier-neutral facilities that are owned and operated by the different leading commercial data center and collocation companies such as Equinix, Telehouse, and InterXion, some of which operate successful for-profit IXPs themselves. The fact that this coming together of such opposing business interests between non-profit and for-profit IXPs does not create frictions or disturbances is easily explained by the fact that the commercial data center companies see great benefit in serving as collocations for these large non-profit IXPs. In fact, when vying for the business of serv-ing as collocations in cities or regions with successful non-profit IXPs, the premise of these commercial data center and collocation companies is that data center or collocation space is more valuable with than without IXP access. These considerations often translate into economic incentives that these companies provide for the "right" IXPs to be in their facilities, especially when these companies themselves provide for-profit Internet exchange services (e.g., the Internet Exchange solution from Equinix or the Global Interlink service offered by Telehouse America).

---

[10] Note that for redundancy purposes, IXPs typically operate two route servers and encourage their members to establish two BGP sessions, one per route server.

[11] In June 2013, AT&T reports carrying 33 petabytes of data traffic on an average business day [4], Deutsche Telekom reports 490 petabytes per month corresponding to 16 petabytes per day on average [12].

| Year | 2003 | 2004 | 2005 | 2006 | 2007 | 2008 | 2009 | 2010 | 2011 | 2012 | 2013 (first half) |
|---|---|---|---|---|---|---|---|---|---|---|---|
| Average Traffic (Gbps) | 13.4 | 24.9 | 58.1 | 110.3 | 193.3 | 288.5 | 443.9 | 623.5 | 815.6 | 1057.0 | 1347.2 |
| Peak Traffic (Gbps) | 22.0 | 47.1 | 119.6 | 220.0 | 374.2 | 608.8 | 856.6 | 1186.0 | 1458.3 | 2042.7 | 2191.0 |
| Number of Members | 178 | 211 | 234 | 253 | 290 | 317 | 349 | 388 | 469 | 555 | 591 |

**Table 1: Annual traffic and member statistics for AMS-IX.**

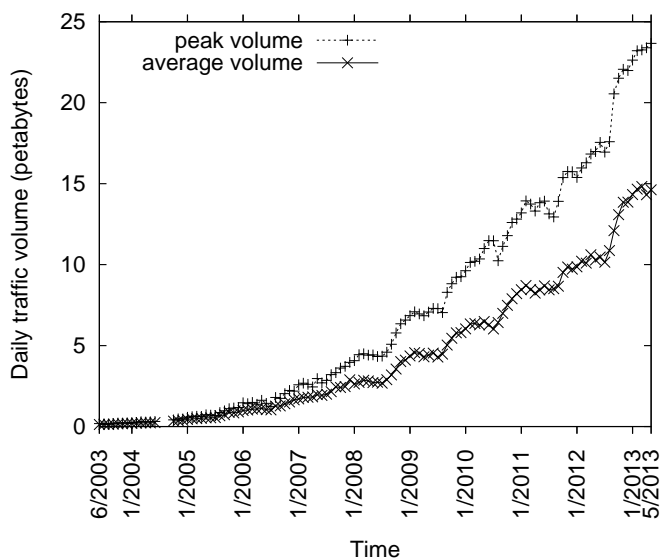

**Figure 4: Daily traffic volumes of AMS-IX in Amsterdam (mid 2003 - mid 2013).**

Third, in view of such often complementary business interests between non-profit and for-profit IXPs, it should come as no surprise to see increasing signs of "joint ventures". In fact, in a recent press release, the managed non-profit DE-CIX announced an expansion of its relationship with the for-profit Equinix by installing new collocations or Points of Presence (PoPs) in three Equinix data centers in Frankfurt [9]. This expanded relationship is mainly driven by DE-CIX's uninterrupted growth in port and data volume demand from its customers and is a first step in transitioning towards DE-CIX's new high-capacity Apollon platform (see Section 2.2). On the one hand, this expansion promises to provide increased collocation capacity for DE-CIX customers, and on the other side, the new collocations will offer enhanced performance and peering opportunities for networks that are Equinix's customers in its Frankfurt data centers. Or, quoting form the press release, "It makes a great deal of sense to implement [DE-CIX's] next-generation architecture in Equinix's network-rich and fast-expanding Frankfurt facilities. Equinix can provide industry-leading collocation with room to grow for [DE-CIX's] members, and the [Equinix's] connectivity-hungry customers can take advantage of [DE-CIX's] very latest peering infrastructure and opportunities." Unfortunately, it seems unlikely that such or similar "joint ventures" between non-profit and for-profit IXPs will provide more visibility into either the customer base of the for-profit IXPs or the traffic that these customers exchange over these IXPs.

## 4. A LOOK AHEAD

For a more research-oriented readership, the following networking-related issues arise naturally from the above discussion about today's Internet IXP marketplace:

- Today's Internet supports a much richer peering fabric than was previously assumed, and this fabric is only expected to get denser in the future.

- As today's largest IXPs continue to experience uninterrupted growth in port and data volume demand from an increasing number of customers, what long-term solutions exist for their hardware (e.g., switching fabric) and software (e.g., increasingly sophisticated service offerings at scale)?

- Obtaining a detailed understanding of the flow of traffic through such an increasingly more densely-connected network poses enormous new challenges.

- A hyper-connected Internet requires a new approach to measurement-driven networking research because it questions the representativeness (and hence usefulness) of most currently available data.

- If recently mentioned intentions to establish the European IXP model with its managed non-profit IXPs in North America materialize and succeed, how would the resulting explosion in new public peering opportunities in North America impact the IXP scene there in particular and the global Internet ecosystem in general?

In the following, we illustrate how acting on some of these issues, observations, and questions could shape entire new research agendas and also connect to a number of networking research topics that are currently considered to be "hot". First, it should be clear by now that all the above-mentioned items follow either directly or indirectly from the single critical observation that public peering in today's Internet is at least an order of magnitude richer than has been assumed in the past – instead of the estimated multiples of tens of thousands of IXP-specific peering links, there are multiples of hundreds of thousands of such links in active use (i.e., carry actual traffic) in the Internet today.

### 4.1 Where IXPs meet SDN

Given this unexpected richness in peering links, it should come as no surprise that there are also indications of a similarly unexpected richness in routing policies. Indeed, in a world that assumes that peering links are scarce and, in turn, AS path diversity is extremely limited, it would be surprising to find that fine-tuned routing policies that perform inter-domain traffic engineering (TE) with BGP are the rule rather than the exception. However, in a network where massive amounts of peering links create unprecedented AS path diversity, inter-domain TE with BGP offers great potential and can be expected to be fully exploited and widely applied. For example, [63] reports encountering routing policies of a connected IXP member at AMS-IX that treat different prefixes of that one and the same member differently and do so depending on the time-of-day, predicted load of the peering link, or other factors. However, what vehicle should be used to facilitate the regular use of such fine-grained routing policies by hundreds of connected networks at a large IXP, most of which may already make use of the IXP's route server?

We have already seen (see Section 3.2) that IXP route servers do not partake in the forwarding path; that is, they do not forward any traffic. This implies a strict separation of routing and forwarding, and in this sense, today's use of route servers in the various IXPs has much in common with Software Defined Networking (SDN). In fact, if SDN can deliver on the promise expressed in [46] "to herald a new day for interdomain routing by allowing BGP's control plane to evolve independently from the underlying switch and router hardware and bringing software control and logic to interdomain routing", then a software defined Internet exchange that builds on the initial concept proposed in [46] may well be the future. This observation is further substantiated by the looming scalability issues for the switching fabrics of the largest IXPs around the world as they experience uninterrupted growth in port and data volume demand from their customers and respond to it by constantly expanding their footprint. However, we posit that ultimately, it will be the IXPs' customer demands for more functionalities at an IXP's route server that will make the case for SDN when it comes to future IXP designs. In fact, there is already high demand for route server functionalities that go beyond standard route filtering, auto-provisioning, or using BGP communities and would require specific support for a wide range of policies for inter-domain TE with BGP, some of which may simply not be possible without SDN.

## 4.2 Where IXPs meet content and the cloud

The observed richness in peering and associated new opportunities for flexible and sophisticated routing policies goes hand-in-hand with a flurry of recent announcements of strategic alliances between ISPs on the one side and CDNs [1, 24, 34, 22] or large content providers on the other side [23, 19, 45]. In practice, these alliances result in the deployment of servers that are owned and operated by these CDNs or large content providers in the ISPs' networks for the sole purpose of serving their content to the end users in those ISPs faster and more efficiently. With most of the CDNs and large content providers being present at and connected to the large IXPs, the ability to efficiently serve content to end users around the world by making good use of these servers deployed in the various third-party networks has appeal, which makes these CDNs and large content providers prime candidates to fully exploit the plethora of new routing policies enabled by the existing rich peering. In turn, accounting of traffic flows within today's Internet and attributing them to the responsible party becomes challenging. For example, an easy-to-pose question like "What is the fraction of today's Internet traffic that is the responsibility of Akamai?" is very difficult to answer when the infrastructures of the various stakeholders overlap more and more and traffic can and is routed in intricate ways that may include public or private peering, traditional customer-provider links, or a combination of all of them; can depend on time-of-day or other factors; and are likely to differ from one network location to the next. Moreover, we fully expect that this situation gets only more challenging with the emergence and rise of cloud providers and their cast of supporting network, many of which are already showing up at the different IXPs.

## 4.3 Where IXPs meet Internet measurement

A hyper-connected Internet where traffic flows in intricate and what often looks like mysterious ways is generally bad news for Internet measurement. In particular, for any set of measurements, the question of representativeness becomes an immediate issue and requires renewed attention by researchers interested in Internet measurement. In particular, questions concerning location, time, and networking conditions need to be revisited when the assumptions made in the past on both the control and data plane are no longer in sync with Internet reality.

At the same time, there is a tremendous upside to Internet measurement with the recent observation that the large European IXPs see traffic from all actively routed ASes, from all countries around the world, and from a large fraction of allocated IPs [49]. As a result, the different IXPs around the world represent a new generation of vantage points with unprecedented visibility into parts of the Internet, where the extent of these parts ranges from highly-confined geographic regions in the case of the small and medium-sized IXPs to the entire world for the largest IXPs. This in turn opens up unique new opportunities for Internet measurement to target cyber-security threats and related issues (e.g., spam, DDoS attacks) that manifest themselves in terms of visible and identifiable traffic on the public peering infrastructure of today's Internet [66]. To this end, the latest blackholing service offering provided by, for example, DE-CIX is only a start, and expanding an IXP's services in this space looms as a promising future direction for both networking researchers and IXPs. Other efforts that are bound to benefit from using this new generation of Internet vantage points include the study of DNS traffic and the analysis of network outages that result from natural disasters, are man-made, or have political causes, to mention but a few.

## 4.4 Where IXPs meet NetEcon

Last but not least, for the more NetEcon-minded networking researchers, given the recently announced intention to establish the "European IXP model" (i.e., managed non-profit IXPs) in the USA under the working title of "Open IX" [67, 52, 40], an all-important question is what such a possible adoption of the European IXP model in the US would do to the North American IXP scene in particular and the worldwide Internet ecosystem in general. Specifically, what are possible scenarios that would make the US Tier-1 ISPs use public peering at a North American IXP when they already do so via their respective ASes in Europe? Similarly, given that the large European IXPs are already in the process of extending their reach globally (e.g., AMS-IX moving into Hong Kong and also into the Caribbean region; DE-CIX operating IXPs also in Hamburg and Munich and since late 2012 also in Dubai, United Arab Emirates), what are possible scenarios that would ensure a successful extension into the North American market? In particular, could the cast of important mobile providers around the world that apparently see benefits in using IXPs for their purpose (e.g., facilitating global data roaming for their mobile end user customers) have enough power to create a more seamless and truly global mobile Internet by encouraging such extensions? Note that while currently many European mobile providers make use of services such AMS-IX GRX (i.e., a full-service, scalable GRX Peering Exchange for the interconnection of mobile GPRS/3G networks within AMS-IX), many of their North American counterparts rely on GPE, an Equinix-specific service that is available for all GRX operators and qualified customers that are collocated within an Equinix IBX and wish to exchange mobile Internet traffic with their peer over the Equinix Internet Exchange. Thus, one question is if this current piecemeal "solution" (i.e., relying on separate IXP-provided ser-vices that are tailor-made for the mobile providers in Europe and North America, respectively) is good enough to result in a status quo where managed non-profit IXPs are unlikely to get a foothold in North America.

## 5. SUMMARY

In the past, IXPs have received little or no attention from the networking researchers who considered them either just as "add-ons" to an Internet dominated by large Tier-1 ISPs and large content providers/distributors or as components of the Internet's infrastructure that are largely void of interesting research problems (after all, an IXP is just a switch ...). We argue in this article that once the research community has a better understanding of the architectures, operations, service offerings, and different business models of the hundreds of operational IXPs in today's Internet, it will likely revise its low opinion of IXPs as a research topic and realize their potential for being a rich source for very interesting and certainly very important problems dealing with a combination of technological and economic issues facing the current and future Internet ecosystem. It is our hope that this article spurs the interest of networking researchers to look into IXPs, entices them to examine in more detail the surveyed literature to learn about the in's and out's of IXPs that this article could not cover, and ultimately convinces the community that there is indeed more to IXPs than meets the (uninformed) eye and that the IXPs when viewed as critical components of the Internet deserve to be taken seriously.

## Acknowledgment

We are very grateful to Arnold Nipper (DE-CIX) and Henk Steenman (AMS-IX) for answering our questions and providing detailed information about their IXPs.

This work was supported in part by the EU projects BigFoot (FP7-ICT-317858) and CHANGE (FP7-ICT-257422), EIT Knowledge and Innovation Communities program, and an IKY-DAAD award (54718944).